# Scientometrics[1]

Loet Leydesdorff [a] and Staša Milojević [b]

[a] Amsterdam School of Communication Research (ASCoR), University of Amsterdam, Kloveniersburgwal 48, 1012 CX Amsterdam, The Netherlands; loet@leydesdorff.net

[b] School of Informatics and Computing, Indiana University, Bloomington 47405-1901, United States; smilojev@indiana.edu.

**Abstract**

The paper provides an overview of the field of scientometrics, that is: the study of science, technology, and innovation from a quantitative perspective. We cover major historical milestones in the development of this specialism from the 1960s to today and discuss its relationship with the sociology of scientific knowledge, the library and information sciences, and science policy issues such as indicator development. The disciplinary organization of scientometrics is analyzed both conceptually and empirically. A state-of-the-art review of five major research threads is provided.

**Cross References:**

Communication: Electronic Networks and Publications; History of Science; Libraries; Networks, Social; Merton, Robert K.; Peer Review and Quality Control; Science and Technology, Social Study of: Computers and Information Technology; Science and Technology Studies: Experts and Expertise; Social network algorithms and software; Statistical Models for Social Networks, Overview;

---

[1] Forthcoming in: Micheal Lynch (Editor), *International Encyclopedia of Social and Behavioral Sciences,* Section 8.5: Science and Technology Studies, Subsection 85030. Elsevier, 2015.

**Introduction**

Scientometrics can be defined as the "quantitative study of science, communication in science, and science policy" (Hess, 1997, 75). What started as Eugene Garfield's idea of an index to improve information retrieval in the 1960s and resulted in the creation of the *Science Citation Index* (SCI) (Garfield, 1979; Wouters, 1999) was soon recognized as a novel instrument in the empirical study of the sciences (e.g., Price, 1965; Cole & Cole, 1973). The availability of output indicators (such as databases of publications and patents) complemented ongoing efforts by the Organization of Economic Cooperation and Development (OECD) in Paris to standardize input statistics of the scientific enterprise (OECD, 1963, 1976). Based on these data, the National Science Board of the U.S.A. initiated the biannual series of *Science Indicators* in 1972.[2]

The new journal *Scientometrics* was launched in 1978 and in that same year leading historians, philosophers of science, and social scientists—among them Robert K. Merton—published an edited volume entitled *Toward a Metric of Science: The Advent of Science Indicators*, in which they reflected on the new perspectives (Elkana et al., 1978). The historian Derek J. de Solla Price published a number of books and articles in the 1960s and '70s which laid the foundations for the newly emerging field of quantitative science studies (e.g., Price, 1961, 1963, 1965), culminating in a full-fledged research program (Price, 1976).

The sociology of science, however, during the 1980s turned increasingly towards micro-analysis focusing on the behavior of scientists in laboratories (e.g., Latour & Woolgar, 1979). From this perspective, the quantitative analysis of scientific literature at the macro (e.g., disciplinary) level was not considered a useful tool to explain scientific practices (Edge, 1979). Rather, with its focus on scientific communications—as a unit of analysis potentially different from scientists as authors—scientometrics developed at arm's length from the sociology of science and closer to the library and information sciences. At the same time, the value of scientometric indicators for informing science policies and research management became manifest (Irvine & Martin, 1984).

---

[2] The series was renamed into *Science and Engineering Indicators* in 1987 (National Science Board, 2012).

Under these diverging pressures, the field of science & technology studies increasingly bifurcated during the period 1985-2000 into qualitative "sociology of scientific knowledge," on the one side, and the quantitative study of scientometrics and science indicators, on the other. Additionally, a third line of research emerged that published articles that use insights from the quantitative study of science and technology for evaluation and policy purposes. Such research appeared in journals such as *Research Policy, Research Evaluation*, *Technology Analysis & Strategic Management* (Leydesdorff & Van den Besselaar, 1997).

During the 2000s, attention to evaluation and ranking was further enhanced after the publication of the first Academic Ranking of World Universities (ARWU) of the Shanghai Jiao Tong University in 2004 (Shin et al., 2011). The use of impact factors of journals for evaluative purposes has in the meantime pervaded the academic environment, even to the level of individual tenure decisions, which increasingly are being based on quantitative measures of publications and citations. Another popular indicator, the *h*-index (Hirsch, 2005), provides a simple impact metric for individual authors that can readily be used in online searching, for example, with Google Scholar, but is also incorporated in the major citation databases such as the Web-of-Science and Scopus. Computer programs (e.g., *Publish or Perish* at http://www.harzing.com/pop.htm) can freely be downloaded from the Internet and allow for measuring numbers of publications, citations, *h*-index, *g*-index (Egghe, 2006), etc., at the level of individuals, journals, and institutions without much prior knowledge of the scientometrics involved. In the meantime, the production and improvement of these indicators has become organized in university departments, spin-offs, and relevant companies such as Elsevier and Thomson-Reuters.

The increased access to large datasets through the Internet led to the development of the network sciences as part of computing and applied physics during the first decade of this century (e.g., Newman, 2010). A number of these studies used coauthorship and citation data for modeling the network dynamics. Although these efforts to model the evolution of the sciences statistically (e.g., Scharnhorst *et al.*, 2012) often do not aim at contributing to social-scientific understanding and theorizing, the new methods (e.g., visualization techniques) developed by these researchers are partly derived from and have also been adopted by scientometricians. Such interdisciplinary exchanges make scientometrics an active research specialty that in the 2000s has been experiencing a spectacular growth in its literature. More recently, the specialty can increasingly be considered as a "research front" in terms of the turnover of the referencing patterns (Milojević & Leydesdorff, 2013; Wouters & Leydesdorff, 1994).

**The disciplinary organization of "scientometrics"**

Unlike the behavioral sciences and mainstream philosophy of science, scientometrics focuses on texts (documents) as empirical units of analysis. Figure 1 schematizes the relations with other disciplinary perspectives in science studies. Texts cannot be reduced to their authors—texts can, for example, be coauthored—nor can theories be reduced to the documents in which they are published. However, a measure in one dimension can be used as a proxy or indicator for the other given a research design.

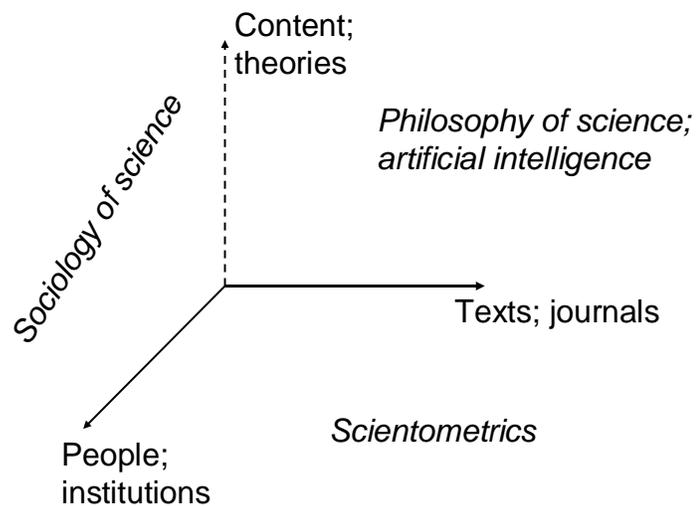

**Figure 1**: Three main dimensions in the dynamics of the sciences; adapted from Leydesdorff (1995).

As against texts or people, contents and theorizing remain latent and thus have to be theorized and hypothesized. Using factor analysis, however, one can reorganize data so that, for example, latent journal structures can be derived from the aggregation of citation linkages among journals. In Figure 2, we applied techniques from network analysis to distinguish four communities in the citation networks among 32 journals that were cited by authors in *Scientometrics* during 2010.[3]

---

[3] All journals are included which contributed more than one percent to the total number of references in the journal *Scientometrics* during this year (2010).

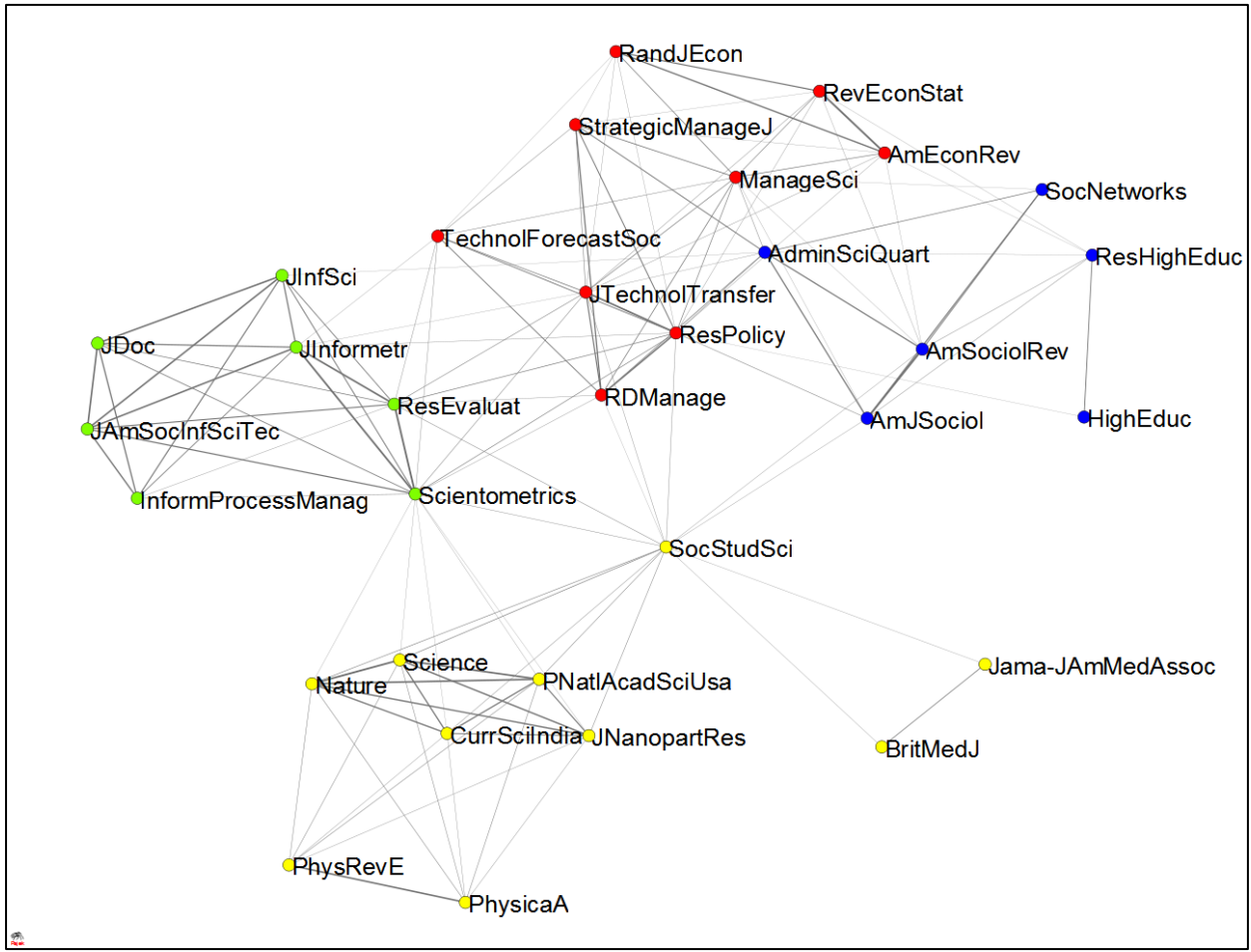

**Figure 2**: Map of 32 journals based on citing documents published in *Scientometrics* during 2010; Pajek used for the visualization; cosine > 0.1; four communities detected with modularity $Q = 0.506$.

Figure 2 shows that *Scientometrics* as a journal has mainly been situated within the information sciences. The *Journal of the American Society for Information Science and Technology* (JASIST) has become a major publication outlet for scholars in this field in addition to the more specialized journals (such as *Journal of Informetrics* that was newly founded in 2007). However, the journal *Scientometrics* is the core journal in this field that interfaces with *Social Studies of Science*; a journal that mainly publishes qualitative science studies, and with general science journals (such as *Science* and *Nature*) and a number of journals that can be considered as part of technology and innovation studies (e.g., *R&D Management* and *Research Policy*). The relation with social network analysis and organization studies is more remote, but also visible.

At the lower-left side of Figure 2, one also can see strong links to a group of major multidisciplinary journals (*Nature, Science, PNAS*) and physics journals (*Physical Review E* and *Physica A*). These journals mostly publish the research of network science researchers—physicists and applied mathematicians—who discovered bibliographic databases as resources and study objects at around the turn of the century. These developments in physics and mathematical statistics have become part of the relevant environments for scientometric research, but the relevant journals are positioned very differently from journals such as *Social Networks* and *Administrative Science Quarterly* (at the top-right side of Figure 2). In other words, scientometric research nowadays is at the crossroads among the social sciences, information science, and advanced computing with its efforts to capture patterns in "big data."

**Major research issues**

*a. The measurement of impact*

The modern citation emerged in scientific literature—notably in chemistry—at the end of the 19th century as a standardized format (Bazerman, 1988; Leydesdorff & Wouters, 1999, 175). Citation indexing was organized during the 1960s (Garfield, 1979; Wouters, 1999; cf. Cronin, 1984). With the advent of Google Scholar and Scopus as alternative citation indexes in 2004, citation analysis and rankings have become increasingly paradigmatic in research management and science policies. Discussions about the meaning of citations and the derived indicators tend to drive the scientometric enterprise.

For several decades, citation impact has been measured as the average number of citations per publication ($c/p$), and the impact factor ($IF$) was defined as a two-year moving average at the level of journals. Citation distributions, however, are right-skewed because of cumulative advantages (Price, 1965, 1976) or, in other words, the "Matthew Effect" in science (Merton, 1968)—the tendency for citation-rich authors and publications to draw further citations, in part *because* they are heavily cited. Therefore, the use of central tendency statistics (such as the arithmetic mean) is ill advised.

In addition to violating statistical assumptions, the use of the mean can affect the calculated impact when applied to different levels of aggregation by placing the number of publications in the denominator. For example, a Principal Investigator (PI) may lose average impact when his/her publications with junior staff are included in the set to be evaluated. However, a research group has more cumulative impact than its individual PI. Obviously, one needs a measure that can be aggregated, but after a normalization of the raw citation scores for differences among fields of science.

In 2010 and 2011, a scholarly debate (e.g., Gingras & Larivière, 2011) led gradually to the acceptance in measuring impact of using percentiles and nonparametric statistics that accord with the skew in the distributions. The *Science & Engineering Indicators* (National Science Board, 2012), for example, uses six classes: the top-1%, top-5%, top-10%, top-25%, top-50%, and

bottom-50%, but these measures can be further developed into appropriate statistics. Percentile ranks remain—like numbers of citations but now normalized—attributes at the article level and can be aggregated at different levels: institutions, nations, or even *œuvres* of individual scholars. Statistical significance can be specified and error indicated, both in comparisons and with reference to expectations. The SCImago Institutions Rankings 2011 and Leiden Rankings 2011 of top-universities, for example, both use the top-10% as an *excellence* indicator for which one can test for differences statistically (Bornmann, de Moya-Anegón & Leydesdorff, 2011).

Statistics based on normalization of the citation curves instead of averaging may lead to very different results. Using 2009 data, for example, Leydesdorff & Bornmann (2011) showed that *Proceedings of the National Academy of Science of the USA* (PNAS) could thus be attributed an impact that was significantly higher than *Science* or *Nature*, although, according to the usual measure, the latter journals had an impact factor that was three times higher. In brief, the new (nonparametric) indicators correlate more strongly with citations and publications than impact factors or other measures based on central tendency statistics.

Hirsch's (2005) *h*-index became another popular impact metric for capturing both the productivity and impact of an individual author. The index itself is crude, but there have been attempts to improve it. For example, Egghe proposed a *g*-index (2006) that increases the sensitivity of *h*-index values to highly cited papers. These indexes can be used for ranking scholars within a single discipline, because publication and citation behavior varies across disciplines. Recently, there have been proposals to create so-called "universal" indexes, that is, indexes that can be used to evaluate and rank scientists across disciplines (e.g., Radicchi et al., 2008).

*b. The delineation of a reference set*

A largely unresolved problem in scientometrics has been the delineation of reference sets for measuring the impact of journals or institutional units. Reference sets are a condition for appropriate normalization. The results of a scientometric evaluation can be highly sensitive to the attribution of papers to particular disciplines or specialties, and to the way specialties and (inter)disciplines are delineated. This makes it very difficult to delineate reference sets for universities that include multi-disciplinary units (Rafols et al., 2012).

Following up on Moed (2010), Leydesdorff and Opthof (2010) proposed to use the citing papers as reference sets and to fractionate the citation credit in accordance with the length of the reference lists. This procedure takes into account that papers can address a variety of audiences with different citation practices. Using all publications in 2005 with Tsinghua University in Beijing as the institutional address, for example, Zhou & Leydesdorff (2011) showed that the department of Chinese Language and Literature would be upgraded from the $19^{th}$ to the $2^{nd}$ position in the ranking among departments of this university if the citation scores were counted fractionally. Although fractionation moderates the effects of differences among disciplines, it does not enable articles that are not cited to be assessed.

In general, the problem remains that the citation matrices can be decomposed into (disciplinary) groupings to a variable extent. A single threshold therefore cannot remove the fuzziness that results from variations among fields of science. At the micro-level, a scientometrics researcher may carefully design appropriate sets of documents for representing relevant environments in considerable detail (Figure 2 is a case in point). However, the specificity in each subset may make it difficult to aggregate sets from which more general conclusions can be drawn. The exception may be the rule in some sciences more than in others.

*c. Theories of citation*

In line with different traditions in the sociology of science, it is possible to distinguish a normative from a constructivist theory of citations and other scientometric artifacts. Following Merton's (1942) set of norms—communalism, universalism, disinterestedness, and organized skepticism—one can argue that citations are a manifestation of influence and can therefore be considered as a currency for paying tribute in science (Zuckerman, 1987). This normative model has been countered by discourse-analytic arguments that emphasize the rhetorical construction of references in texts (Cozzens, 1989; Gilbert, 1977).

Fujigaki (1998) provided an interpretation of referencing in scientific papers as an important part of the self-organizing system in scientific fields: by providing citations, authors select those elements from the previously existing knowledge base which have to be retained for further development in a next round. Leydesdorff (1998) added that texts and authors constitute layered networks in which citations are expected to have different meanings. The system in the present can thus move forward referencing both the social and intellectual organization of science, while potentially following trajectories based on mutual adjustments between selections in these dimensions. Trajectories, however, can also be meta-stabilized and then globalized into regimes or paradigms. An evolutionary model of science is thus envisaged that is open for simulation and empirical research (Scharnhorst et al., 2012).

*d. Mapping science*

Following Rip (1988, 254), a map of science can be defined as "the visualization of the topology of relationships between elements or aspects of science." Mapping of science can be useful for three purposes: retrieving information, understanding the dynamics of science, and informing science-policy decisions about the allocation of resources (i.e., funding) and rewards. Scientometric analysis focuses on revealing the internal structure of intellectual domains, that is, mapping the components of disciplines, fields, or specialties on the basis of evidence from the literatures under study. This can be achieved by mapping subject terms, documents, authors' *œuvres*, or journals. The basic data are co-occurrence counts.

Callon et al. (1986) argued that the sciences develop as heterogeneous networks. In addition to heterogeneity across institutional domains (university, industry, government), organized knowledge can flexibly be codified in terms of mixtures of cited references and word usages. Words are more volatile than citations. Coauthorship relations, however, are not sufficiently informative of intellectual content for a semantic map; social network analysis has to be combined with semantic maps in order to generate rich representations. A longer-term challenge for this agenda is the animation of the informed networks over time, preferentially with specification of the statistical error involved. Advances in visualization and animation can further inform our understanding of the role of citations, title-words, keywords, author names, etc., in evolving networks of scientific communication.

*e. Policy and management contexts*

Scientometricians also construct indicators that can be used in policy and management contexts. In these contexts, utility is sometimes more important than validity. For example, the use of impact factors to evaluate faculty members involves a combination of analyses at different levels of aggregation, and eventual applications to relatively small sets. Science maps and overlays to geographical maps can function heuristically at the aggregate level, but testing for significance requires sufficiently large sets.

The use of scientometric indicators for strategic choices at low levels of aggregation, such as "picking the winners" (Irvine & Martin, 1984) in competitions for funding, is sensitive to the problem of relatively small sets, the operationalization of quality (e.g., the normalization of indicators across specialties and disciplines), and also the intrinsic problem of selecting the "excellent" candidates for funding from an already pre-selected set of fundable proposals. Bornmann *et al.* (2010) showed in a number of cases that "best rejected" proposers, when matched as pairs with grantees, scored higher on performance and impact indicators at a statistically significant level. The peer review process may introduce normative bias without intending to do so, in a process of selecting "excellent" proposals from a pool of "good" proposals. Scientometric meta-evaluation may help with improving the selection process by enhancing awareness that it is not always possible to distinguish "excellent" from "good" research despite pressures for doing so for policy reasons.

**Conclusions and further perspectives**

The scientometric perspective adds a quantitative focus on texts and communication to the interdisciplinarity of science and technology studies. The philosophical distinction between "context of discovery" and "context of justification" (Popper, [1935], 1959) was first overcome in empirical science studies that focus on interactions in which the social and cognitive organization of the sciences are continuously and actively reconstructed and recombined. However, this social and intellectual process is textually mediated because texts can "travel" more easily than scholars do, and thus the global dimension of paradigmatically structured horizons of meaning can be instantiated in relation to locally generated novelty. At the level of texts (e.g., manuscripts), one is able to recombine references to both local agency (author names, institutional addresses) and cognitive organizers such as title words and journal names. Furthermore, the content of manuscripts is validated (by peer review in contexts of justification) and thereupon the texts are admitted to the archive of published, and thus authenticated, scholarship on which future work can be built. These processes can be traced because they are documented.

The field of scientometrics has also expanded to different types of documents and other domains. For example, a similar process is working in the technological domain, but with a different dynamics (Dosi, 1982). Like manuscripts, patent applications contain knowledge claims that refer to "prior art". However, the functions and therefore the institutional incentives are different: patents are meant to protect intellectual property, whereas the public sciences are also based on the principle of gift-giving (Merton, 1973). As publications, patents can be searched on the internet, and the data can then be reorganized in terms of lines of intellectual heritage using citations, inventor names, institutional addresses, etc. Overlays on Google Maps, for example, allow for studying the geographic diffusion of new technologies based on sets and series of patented inventions.

Texts span networks of relations among authors, inventors, cognitions, institutional addresses, journals, etc. These networks contain both social (including economic) and cognitive relations. The cognitive dynamics are different from the social dynamics in that meaning is reflexively provided (that is, from the perspective of hindsight), while the historical networks develop with the time arrow. Thus, forward and backward loops are intertwined. Whereas the forward arrow necessarily generate Shannon-type information—uncertainty, variation—the backward loops can selectively reduce uncertainty by providing meanings to the events. Using the theory and computation of anticipatory systems (Dubois, 1998), the two operations can be distinguished as reconstructive incursion versus historical recursion.

In summary, the modeling of knowledge exchanges in scientific discourses cannot be reduced to the exchanges of information in co-authorship, co-word, or citation relations. Models as entertained in the sciences enable researchers both to provide meaning to possible future states and to specify uncertainty. The measurement of the communication/sharing of meaning among frames of reference—for example, in university-industry-government relations—is very much on the research agenda of scientometrics (e.g., Leydesdorff & Ivanova, in press), but still a step away from how meaningful communication can further be codified in scientific discourses.

Models also serve to communicate possible future states. A construction of future states from a knowledge-based perspective can be modeled as hyper-incursion. The modelers/scientists thus become carriers who are differently positioned in terms of their reflexive and communicative competencies; for example, as scientific explorers and/or appliers of engineering knowledge. The (re)constructions and their interactions update and reinforce the knowledge bases of the evolving societies and their economies. Authors in scientometrics are able to contribute to the study of science, technology, and innovation from a quantitative perspective by modeling and measuring these developments.